\documentclass{revtex4}

\usepackage{graphicx}
\usepackage{mathtools}
\usepackage{ulem}
\usepackage{color}
\usepackage{amsmath}
\usepackage{amssymb}
\usepackage{accents}
\usepackage{pstricks,pst-node,pst-coil,pst-plot}
\usepackage{dcolumn}
\usepackage{bm}
\usepackage{mathrsfs}
\usepackage{slashed}
\usepackage{epstopdf}
\usepackage[mathcal]{euscript}
\usepackage{url}
\usepackage{hyperref}  

\def\bL{\boldsymbol{L}}

\def\bQ{\boldsymbol{Q}}
\def\bep{\boldsymbol{\epsilon}}

\def\d{\mathrm{d}}

\allowdisplaybreaks

\allowdisplaybreaks
\begin{document}
\title{The first law of black hole mechanics in conformal Einstein-Power-Yang-Mills theory}

\author{Xiaokai He}\email[Xiaokai He: ]{sjyhexiaokai@hnfnu.edu.cn}
\affiliation{School of Mathematics and Statistics\\
 and Hunan Provincial University Key Laboratory for Big Data Analysis and Application, Hunan First Normal University, Changsha 410205, China}

\author{Xiaoning Wu
\footnote{corresponding author}
} \email[Xiaoning Wu: ]{wuxn@amss.ac.cn}
\affiliation{Institute of Mathematics, Academy of Mathematics and Systems Science and State Key Laboratory of
Mathematical Sciences, Chinese Academy of Sciences, Beijing 100190, China}
\affiliation{ School of Mathematical Sciences, University of Chinese Academy of Sciences, Beijing 100049, China}

\author{Naqing Xie} \email[Naqing Xie: ]{nqxie@fudan.edu.cn}
\affiliation{School of Mathematical Sciences, Fudan
University, Shanghai 200433, China}

\begin{abstract}
In the framework of the Iyer–Wald formalism, the first law of black hole mechanics is examined within the context of conformal Einstein–power–Yang–Mills (CEPYM) theory. By comparing two infinitesimally neighbouring stationary black hole solutions, we obtain the explicit analytical expression for the first law of black hole thermodynamics in CEPYM theory.
\end{abstract}

\maketitle

\section{Introduction}

The landmark work  of the four laws of black hole mechanics constitutes a fundamental achievement in the theoretical investigation of black holes \cite{BCH1973}. Within the framework of Einstein’s general relativity, the first law of black hole mechanics governs the first-order variations of the mass $M$, horizon area $A$, angular momentum $J$, and electric charge $Q$ for a generic stationary, charged, and rotating black hole, satisfying the relation
\begin{eqnarray}\label{1stlaw-00}
\delta M=\frac{\kappa}{8\pi}\delta A+\Omega_H\delta J+\Phi\delta Q,
\end{eqnarray}
where $\kappa$ is the surface gravity of the stationary black hole, $\Omega_H$ and $\Phi$ are the angular velocity and the electric potential of the horizon respectively.
The original formulation of the first law, due to Bardeen, Carter and Hawking, differs slightly from eq. (\ref{1stlaw-00}), since their analysis in Ref.~\cite{BCH1973} considered only a perfect fluid in circular orbits around the black hole.

The covariant phase space approach, pioneered by Iyer and Wald in 1994 \cite{IW1994}, has provided a generic framework for deriving the first law of black hole mechanics. 
Subsequent developments have extended this formalism, with widespread applications to black hole thermodynamics and conserved-charge analyses in asymptotically flat spacetimes \cite{Zoupas2000,Gao2003,Fang-He2018,Jiangjie2020,Jiangjie2021,Grant2022,Xiaoyong2024,Z-hongbao2024,Z-hongbao2025,HWX2026}. In their original derivation, Iyer and Wald crucially exploited the geometric property of the bifurcation two-sphere, on which the horizon Killing vector vanishes identically. Specifically, they adopted the bifurcation sphere as a boundary component of a spacelike hypersurface that extends all the way to spatial infinity. A notable generalization came in 2003, when Gao extended the original formulation of Ref. ~\cite{IW1994} to generic charged rotating black hole configurations
in which the  fields are not required to be smooth across the event horizon. This generalization was achieved by replacing the bifurcation sphere with an arbitrary spatial cross-section of the event horizon which then serves as the boundary of the underlying spacelike hypersurface \cite{Gao2003}. 
More explicitly, Gao rigorously derived the first law of black hole mechanics for both Einstein–Maxwell and Einstein–Yang–Mills theories.

Classical general relativity predicts that black hole spacetimes generically host curvature singularities, which may indicate that the classical description of gravity breaks
down in the strong-field regime. As early as 1968, Bardeen constructed an explicit regular black hole spacetime devoid of a central curvature singularity \cite{Bardeen1968}. The so-called Bardeen solution The so-called Bardeen solution stands as a prototype of regular black holes, and have since garnered sustained research interest across the community. Comprehensive review articles on this subject are available in Refs. \cite{Lemos2011,Miao2023}.  Within the framework of Einstein gravity minimally coupled to nonlinear electrodynamics, the source of Bardeen solution can be interpreted as a  magnetic monopole \cite{Beato1998,Beato2000}.   More recently, in 2018, Zhang and Gao investigated the first law and Smarr formula of black hole mechanics  in nonlinear Abelian gauge theories\cite{Gao2018}.

As one of the most fundamental non-Abelian gauge theories, Yang–Mills theory has achieved unprecedented success in describing fundamental particle interactions \cite{Peskin1995}. Correspondingly, gravitational systems coupled to Yang–Mills fields have been widely investigated over the past several decades \cite{BM1988,MH2008}. Motivated by these developments, the present paper aims to derive  the first law of black hole mechanics within the framework of conformally invariant Einstein–Yang–Mills coupled gravity.


To construct a conformally invariant gravitational theory in $D$ dimensions, it is necessary to introduce power-type Yang–Mills invariants.  The Einstein-power–Yang–Mills (EPYM) action was proposed in 2009 \cite{HH2009}:
\begin{eqnarray}
I_{\rm{EPYM}}=\frac{1}{2}\int\d^Dx\sqrt{-g}[R-(\mathcal{F}_{YM})^q],
\end{eqnarray}
which reduces to the standard Einstein–Yang–Mills action when $q=1$. 
In particular, the EPYM action is conformally invariant for $D=4q$.


In 2021, Li and Miao generalized the Einstein-power– Yang–Mills theory to a conformally invariant gravitational framework, namely the conformal Einstein-power–Yang–Mills (CEPYM) theory, whose action takes the form \cite{Li-Miao2021}
\begin{eqnarray}
&&\ \ \ I_{\rm{CEPYM}}\nonumber\\
&&=\frac{1}{2}\int\d^Dx\sqrt{-g}\bigg(
\frac{1}{4}\frac{D-2}{D-1}R\phi^2-\phi\Box\phi-(\mathcal{F}_{\rm YM})^q\bigg).\ \ 
\end{eqnarray}
When $D=4q$, the CEPYM action is invariant under the conformal transformations
\begin{eqnarray}
\hat{g}_{\mu\nu}=\Omega^2(x)g_{\mu\nu},\\
\hat{\phi}=\Omega^{-1}(x)\phi,\\
\hat{A}^a_{\mu}=A^a_{\mu},
\end{eqnarray}
where $\Omega(x)$ is an arbitrary smooth positive function.

In Ref.~\cite{Li-Miao2021}, the authors studied the thermodynamics and dynamics of high-dimensional Einstein-power-Yang–Mills black holes in conformal gravity. The CEPYM theory admits a regular black hole solution, whose Kretschmann scalar is finite as $r$ goes to zero.  Assuming the first law of black hole thermodynamics, they obtained the entropy expression \cite[Eqs. (44)-(45)]{Li-Miao2021}. In the present paper, we aim to establish the first law of black hole mechanics for the CEPYM theory from first principles. Using the covariant phase space formalism, we derive an explicit analytical expression for the first law of black hole thermodynamics in four-dimensional CEPYM gravity.

The remainder of this paper is organized as follows.  In Sec.\ref{sec2}, beginning with the Lagrangian 4-form of the 4-dimensional CEPYM theory, we derived the symplectic potential $\boldsymbol{\Theta}$, the Noether current 3-form $\boldsymbol{J}_{\xi}$ associated with a vector field $\xi$, and the  corresponding Noether charge 2-form $\boldsymbol{Q}_{\xi}$.   In Sec.\ref{sec3}, we obtain the explicit form of the first law of black hole mechanics for CEPYM gravity. Finally, conclusions are presented in Sec.\ref{conclusion}.
 
Throughout this paper, we use a geometric unit system with the speed of light and gravitational constant taken to be one ($c=G=1$).

\section{Covariant phase space formalism for 4-dimensional CEPYM theory}\label{sec2}

The Lagrangian 4-form of the 4-dimensional CEPYM theory reads \cite{Li-Miao2021}
\begin{eqnarray}
 \boldsymbol{L}=\frac{1}{2}\bigg(\frac{1}{6}R\phi^2-\phi g^{\mu\nu}\nabla_{\mu}\nabla_{\nu}\phi -\mathcal{F}_{\rm YM}\bigg)\boldsymbol{\epsilon},
\end{eqnarray}
where $\boldsymbol{\epsilon}$ is the spacetime volume form and
\begin{eqnarray}
&&\mathcal{F}_{\rm YM}=\sum_{a=1}^3g^{\mu\nu}g^{\alpha\beta}F_{\mu\alpha}^aF_{\nu\beta a}=F^a_{\nu\beta} F_{a\nu\beta},\\
    &&F^{a}_{\mu\nu}
=\partial_{\mu}A_{\nu}^a-\partial_{\nu}A_{\mu}^a+\frac{1}{2\sigma}C^a_{ \ bc}A^b_{\mu}A^c_{\nu}.
\end{eqnarray}
Here $A_{\mu}^a$ are the gauge potentials, $C_{\ bc}^a$ and $\sigma$ denote the structure constants of SO(3) Lie group and coupling constant, respectively. Here we use the Latin indices, $a,b,\cdots=1,2,3$ to represent the internal space of the gauge group, which are raised and lowered with respect to the Killing form of the Lie algebra $so(3)$. The Greek letters $\mu,\nu,\cdots=0,1,2,3$ are used as spacetime indices.

\subsection{The symplectic potential 3-form}

Within the covariant phase space formalism, the derivation of the symplectic potential relies critically on computing the first-order variation of the gravitational Lagrangian 4-form associated with the CEPYM theory. To proceed, we first decompose the full Lagrangian density
\begin{eqnarray}\label{lagrangeL}
   L=\frac{1}{2}\bigg(\frac{1}{6}R\phi^2-\phi g^{\mu\nu}\nabla_{\mu}\nabla_{\nu}\phi -\mathcal{F}_{\rm YM}\bigg)
\end{eqnarray}
into three  sectors corresponding to the conformal gravity scalar-curvature contribution, the auxiliary scalar field kinetic term, and the Yang–Mills field interaction term:
\begin{eqnarray}
&&L=L^g+L^{\phi}+L^{\rm{YM}},
\end{eqnarray}
with
\begin{eqnarray}
    &&L^g=\frac{1}{12}R\phi^2,\\
    && L^{\phi}=-\frac{1}{2}\phi g^{\mu\nu}\nabla_{\mu}\nabla_{\nu}\phi,\\
    &&L^{\rm{YM}}=-\frac{1}{2}F_{\mu\nu}^aF^{a\mu\nu }.
\end{eqnarray}
Then 
\begin{eqnarray}
&&\delta\bL=(\delta L)\bep+L\delta\bep\nonumber\\
&&\ \ =(\delta L^g+\delta L^{\phi}+\delta L^{\rm{YM}})\bep-\frac{1}{2}L\bep g_{\mu\nu}\delta g^{\mu\nu}.
\end{eqnarray}

We first evaluate $\delta L^g$. Making use of the well-known  variation formula for the Ricci scalar
\begin{eqnarray}
    \delta R=\nabla^{\mu}\bigg(
    g^{\alpha\beta}(\nabla_{\beta}\delta g_{\mu\alpha}-\nabla_{\mu}\delta g_{\alpha\beta})
    \bigg)+R_{\mu\nu}\delta g^{\mu\nu},
\end{eqnarray} 
one can get
\begin{eqnarray}
 &&\ \ \  \frac{1}{12}\phi^2\delta R\nonumber\\
 & &=   \frac{1}{12}\phi^2\nabla^{\mu}\bigg(
    g^{\alpha\beta}(\nabla_{\beta}\delta g_{\mu\alpha}-\nabla_{\mu}\delta g_{\alpha\beta})
    \bigg)+ \frac{1}{12}\phi^2R_{\mu\nu}\delta g^{\mu\nu}\nonumber\\
&&=\nabla_{\mu}\bigg[\frac{1}{12}\phi^2g^{\mu\nu}g^{\alpha\beta}(\nabla_{\beta}\delta g_{\nu\alpha}-\nabla_{\nu}\delta g_{\alpha\beta})\nonumber\\
&&\ \ \ -\frac{1}{6}\phi(\nabla^{\nu}\phi)g^{\alpha\mu}\delta g_{\nu\alpha}+\frac{1}{6}\phi(\nabla^{\mu}\phi)g^{\alpha\beta}\delta g_{\alpha\beta}\bigg]\nonumber\\
&&\ \ \ +\frac{1}{12}\bigg[\phi^2R_{\mu\alpha}-2(\nabla_{\alpha}\phi)\nabla_{\mu}\phi-2\phi\nabla_{\alpha}\nabla_{\mu}\phi\nonumber\\
&&\ \ \ +2(\nabla_{\beta}\phi)(\nabla^{\beta}\phi)g_{\mu\alpha}
+2\phi(\nabla_{\beta}\nabla^{\beta}\phi)g_{\mu\alpha}\bigg]\delta g^{\mu\alpha}.
\end{eqnarray}
Therefore,
\begin{equation}\label{deltalg}
\begin{split}
\delta L^g=&\frac{1}{12}\phi^2\delta R+\frac{1}{6}R\phi\delta\phi\\
=&\nabla_{\mu}\bigg[\frac{1}{12}\phi^2g^{\mu\nu}g^{\alpha\beta}(\nabla_{\beta}\delta g_{\nu\alpha}-\nabla_{\nu}\delta g_{\alpha\beta})\\
& -\frac{1}{6}\phi(\nabla^{\nu}\phi)g^{\alpha\mu}\delta g_{\nu\alpha}+\frac{1}{6}\phi(\nabla^{\mu}\phi)g^{\alpha\beta}\delta g_{\alpha\beta}\bigg]\\
& +\frac{1}{12}\bigg[\phi^2R_{\mu\alpha}-2(\nabla_{\alpha}\phi)\nabla_{\mu}\phi-2\phi\nabla_{\alpha}\nabla_{\mu}\phi\\
& +2(\nabla_{\beta}\phi)(\nabla^{\beta}\phi)g_{\mu\alpha}
+2\phi(\nabla_{\beta}\nabla^{\beta}\phi)g_{\mu\alpha}\bigg]\delta g^{\mu\alpha}\\
&+\frac{1}{6}R\phi\delta\phi.
\end{split}
\end{equation}

Next we compute the variation of the scalar-field Lagrangian $L^{\phi}$. Direct calculation yields 
\begin{eqnarray}
&&\delta L^{\phi}=-\frac{1}{2}g^{\mu\nu}(\nabla_{\mu}\nabla_{\nu}\phi)\delta\phi
-\frac{1}{2}\phi(\nabla_{\mu}\nabla_{\nu}\phi)\delta g^{\mu\nu}\nonumber\\
&&\ \ \ \ \ \ \  \ \ -\frac{1}{2}\phi g^{\mu\nu}\delta(\nabla_{\mu}\nabla_{\nu}\phi).
\end{eqnarray}
Using the following two fundamental variation identities
\begin{eqnarray}
\delta(\nabla_{\mu}\nabla_{\nu}\phi)
&=&\nabla_{\mu}\nabla_{\nu}\delta\phi-(\delta\Gamma^{\alpha}_{\mu\nu})\nabla_{\alpha}\phi
\end{eqnarray}
and
\begin{eqnarray}
\delta\Gamma^{\alpha}_{\mu\nu}=\frac{1}{2}g^{\alpha\beta}(\nabla_{\mu}\delta g_{\nu\beta}+\nabla_{\nu}\delta g_{\mu\beta}-\nabla_{\beta}\delta g_{\mu\nu}),
\end{eqnarray}
one can obtain
\begin{equation}\label{deltalphi}
\begin{split}
\delta L^{\phi}&= \nabla_{\mu}\bigg[-\frac{\phi}{2}g^{\mu\nu}\nabla_{\nu}\delta\phi+\frac{1}{2}g^{\mu\nu}(\nabla_{\nu}\phi)\delta\phi\\
& +\frac{1}{2}g^{\mu\nu}g^{\alpha\beta}(\phi\nabla_{\alpha}\phi)\delta g_{\nu\beta}
-\frac{1}{4}g^{\beta\nu}g^{\alpha\mu}(\phi\nabla_{\alpha}\phi)\delta g_{\beta\nu}\bigg]\\
&+\bigg[
\frac{1}{2}(\nabla_{\mu}\phi)(\nabla_{\nu}\phi)
-\frac{1}{4}g^{\alpha\beta}(\nabla_{\alpha}\phi)(\nabla_{\beta}\phi)g_{\mu\nu}
\\
&-\frac{\phi}{4}(\nabla_{\alpha}\nabla^{\alpha}\phi)g_{\mu\nu}
\bigg]\delta g^{\mu\nu}-g^{\mu\nu}(\nabla_{\mu}\nabla_{\nu}\phi)\delta\phi.\nonumber
\end{split}
\end{equation}

We now proceed to compute $\delta L^{\rm{YM}}$. It contains two distinct classes of contributions which are the variations induced by metric perturbation $\delta g^{\mu\nu}$ and those originating from infinitesimal shifts of the gauge potential $\delta A^c_{\alpha}$. It follows the definition of $L^{\rm{YM}}$ that

\begin{eqnarray}
&&\ \ \ \delta L^{\rm{YM}}\nonumber\\
&&= -\frac{1}{2}g^{\alpha\beta}F^a_{\mu\alpha}F_{\nu\beta a}\delta g^{\mu\nu}
-\frac{1}{2}g^{\mu\nu}F^a_{\mu\alpha}F_{\nu\beta a}\delta g^{\alpha\beta}\nonumber\\
&&\ \ \  -\frac{1}{2}g^{\mu\nu}g^{\alpha\beta}(\delta F^a_{\mu\alpha})F_{\nu\beta a}-\frac{1}{2}g^{\mu\nu}g^{\alpha\beta}F^a_{\mu\alpha}\delta F_{\nu\beta a}.
\end{eqnarray}

By using the relation
\begin{eqnarray}
&&\delta F^{a}_{\mu\alpha}=\delta(\partial_{\mu}A^a_{\alpha}
-\partial_{\alpha}A^a_{\mu}+\frac{1}{2\sigma}C^a_{bc}A^b_{\mu}A^c_{\alpha})\nonumber\\
&&\ \ \ \ \ \ \ =
\partial_{\mu}\delta A^a_{\alpha}-\partial_{\alpha}\delta A^a_{\mu}+\frac{1}{2\sigma}C^a_{bc}(\delta A^b_{\mu})A^c_{\alpha}\nonumber\\
&&\ \ \ \ \ \ \ \ \ \ +\frac{1}{2\sigma}C^a_{bc} A^b_{\mu}(\delta A^c_{\alpha})\nonumber\\
&&\ \ \ \ \ \ \ =
\nabla_{\mu}\delta A^a_{\alpha}-\nabla_{\alpha}\delta A^a_{\mu}+\frac{1}{2\sigma}C^a_{bc}(\delta A^b_{\mu})A^c_{\alpha}\nonumber\\
&&\ \ \ \ \ \ \ \ \ \ +\frac{1}{2\sigma}C^a_{bc} A^b_{\mu}(\delta A^c_{\alpha}),\ \ 
\end{eqnarray}
one can obtain
\begin{eqnarray}\label{deltalym}
&&\delta L^{\rm{YM}}
= -2
\nabla_{\mu}(F^{\mu\alpha a}\delta A_{\alpha a})-g^{\alpha\beta}F^a_{\mu\alpha}F_{\nu\beta a}\delta g^{\mu\nu}\nonumber\\
&&\ \ \ \ \ \ \ \ \ \ \ \ +
 2\bigg(
 \nabla_{\mu}F^{\mu\alpha 
 }_a+\frac{1}{2\sigma}F^{\mu\alpha b}C_{bac}A^c_{\mu}
\bigg )\delta A^a_{\alpha},
\end{eqnarray}
where we have used the fact
\begin{eqnarray}
F^a_{\mu\nu}=-F^a_{\nu\mu},\ \ C_{ \ bc}^a=-C^a_{ \ cb}.
\end{eqnarray}

Summing the variational contributions from all three sectors and substituting back into the full 4-form variation, we reorganize $\delta\bL$ into the sum of a total exterior derivative and the terms proportional to the independent field variations $\delta g^{\mu\nu},\delta\phi$ and $\delta A^a_{\alpha}$ as follows:
\begin{eqnarray}
&&\ \ \ \delta\bL\nonumber\\
&&=(\delta L^g+\delta L^{\phi}+\delta L^{\rm{YM}})\bep-\frac{1}{2}\bep L g_{\mu\nu}\delta g^{\mu\nu}\nonumber\\
&&=\boldsymbol{\epsilon}\nabla_{\mu}\theta^{\mu} +\boldsymbol{E}^g\delta g^{\mu\nu}+\boldsymbol{E}^{\phi}\delta\phi+\boldsymbol{E}^{\rm{YM}}\delta A^a_{\alpha},
\end{eqnarray}
where
\begin{eqnarray}
\theta^{\mu}&=&
\frac{1}{12}\phi^2g^{\mu\nu}g^{\alpha\beta}(\nabla_{\beta}\delta g_{\nu\alpha}-\nabla_{\nu}\delta g_{\alpha\beta})\nonumber\\
&&+\frac{1}{6}\phi(\nabla^{\nu}\phi)g^{\beta\mu}\delta g_{\nu\beta}+\frac{1}{6}\phi(\nabla^{\beta}\phi)g^{\nu\mu}\delta g_{\nu\beta}\nonumber\\
&&-\frac{1}{12}\phi(\nabla^{\mu}\phi)g^{\alpha\beta}\delta g_{\alpha\beta}
-\frac{1}{2}\phi g^{\mu\nu}\nabla_{\nu}\delta\phi\nonumber\\
&&+\frac{1}{2}g^{\mu\nu}(\nabla_{\nu}\phi)\delta\phi
-2F^{\mu\alpha a}\delta A_{\alpha a},
\end{eqnarray}
and
\begin{eqnarray}
&&\boldsymbol{E}^g=\boldsymbol{\epsilon}\bigg[\frac{1}{12}\phi^2\big(R_{\mu\nu}-\frac{1}{2}Rg_{\mu\nu}\big)+\frac{1}{3}(\nabla_{\mu}\phi)(\nabla_{\nu}\phi)\nonumber\\
&&\ \ \ \ \ \ \ -\frac{1}{6}\phi\nabla_{\mu}\nabla_{\nu}\phi-\frac{1}{12}(\nabla_{\beta}\phi)(\nabla^{\beta}\phi)g_{\mu\nu}\nonumber\\
&&\ \ \ \ \ \ \ +\frac{1}{6}\phi(\nabla_{\beta}\nabla^{\beta}\phi)g_{\mu\nu}-g^{\alpha\beta}F^a_{\mu\alpha}F_{\nu\beta a}\nonumber\\
&&\ \ \ \ \ \ \ +\frac{1}{4}F^{\alpha\beta a}F_{\alpha\beta a}
g_{\mu\nu}\bigg],\label{E-28}\\
&&\boldsymbol{E}^{\phi}=\boldsymbol{\epsilon}\bigg[\frac{1}{6}R\phi-\nabla_{\beta}\nabla^{\beta}\phi\bigg],\label{E-29}\\
&&\boldsymbol{E}^{\rm{YM}}=\boldsymbol{\epsilon}\bigg[2\nabla_{\mu}F^{\mu\alpha}_a+\frac{1}{\sigma}F^{\mu\alpha b}C_{bac}A^c_{\mu}\bigg].\label{E-30}
\end{eqnarray}
Finally, using the identity 
$$\boldsymbol{\epsilon}\nabla_{\mu}\theta^{\mu}=d(\theta\cdot\boldsymbol{\epsilon})$$
 for the Hodge dual contraction between the vector $\theta^{\mu}$
 and the volume 4-form, we rewrite the total Lagrangian variation in the standard covariant-phase-space canonical form
\begin{eqnarray}
\delta\boldsymbol{L}=\boldsymbol{E}^g\delta g^{\mu\nu}+\boldsymbol{E}^{\phi}\delta \phi+\boldsymbol{E}^{\rm{YM}}\delta A^a_{\alpha}+\d \boldsymbol{\Theta},
\end{eqnarray}
where
\begin{eqnarray}
\boldsymbol{\Theta}=\theta\cdot\bep
\end{eqnarray}
is the so-called symplectic potential 3-form, with  $\xi\cdot\boldsymbol{\epsilon}$  the contraction of $\theta^{\mu}$ with the first index of $\boldsymbol{\epsilon}$, i.e.,
$$(\theta\cdot\boldsymbol{\epsilon})_{\nu\alpha\beta}=\theta^{\mu}\epsilon_{\mu\nu\alpha\beta}.$$

Defining
\begin{eqnarray}
&&(\boldsymbol{\Theta}^g)_{\alpha_2\alpha_3\alpha_4}=\frac{1}{12}\epsilon_{\mu\alpha_2\alpha_3\alpha_4}\phi^2g^{\mu\nu}g^{\alpha\beta}\big(\nabla_{\beta}\delta g_{\nu\alpha}\nonumber\\
&&\ \ \ \ \ \ \ \ \ \ \ \ \ \ \ \ \ \ \ -\nabla_{\nu}\delta g_{\alpha\beta}\big),\\
&&(\boldsymbol{\Theta}^{\phi})_{\alpha_2\alpha_3\alpha_4}
=\epsilon_{\mu\alpha_2\alpha_3\alpha_4}\bigg(\frac{1}{3}\phi(\nabla^{\nu}\phi)g^{\beta\mu}\delta g_{\nu\beta}\nonumber\\
&&\ \ \ \ \ \ \ \ \ \ \ \ \ \ \ \ \ \ \ -\frac{1}{12}\phi(\nabla^{\mu}\phi)g^{\alpha\beta}\delta g_{\alpha\beta}\nonumber\\
&&\ \  \ \ \ \ \ \ \ \ \ \ \ \ \ \ \ \ \ 
-\frac{1}{2}\phi g^{\mu\nu}\nabla_{\nu}\delta\phi+\frac{1}{2}g^{\mu\nu}(\nabla_{\nu}\phi)\delta\phi\bigg),\ \ \\
&&(\boldsymbol{\Theta}^{\rm{YM}})_{\alpha_2\alpha_3\alpha_4}=-2\epsilon_{\mu\alpha_2\alpha_3\alpha_4}F^{\mu\alpha a}\delta A_{\alpha a},
\end{eqnarray}
we have that the full symplectic potential can be decomposed into
\begin{eqnarray}
   \boldsymbol{\Theta}= \boldsymbol{\Theta}^{g}+\boldsymbol{\Theta}^{\phi}+\boldsymbol{\Theta}^{\rm{YM}}.
\end{eqnarray}

\subsection{Noether current 3-form}

The Noether current 3-form associated with a  vector field $\xi^{\mu}$ is defined by \cite{IW1994}
\begin{eqnarray}
\boldsymbol{J}_{\xi}=\boldsymbol{\Theta}(\tilde{\phi},\mathcal{L}_{\xi}\tilde{\phi})-\xi\cdot\boldsymbol{L},
\end{eqnarray}
where $\tilde{\phi}=(g_{\mu\nu},\phi,A_{\mu}^a)$ denotes the collection of the dynamical fields. In this subsection, we will show that the Noether current 3-form can be split into a constraint 3-form and the exterior differentiation of the Noether charge 2-form, and give the explicit expressions of the Noether charge 2-form.

Upon substituting the explicit form of the symplectic potential $\boldsymbol{\Theta}$ and the Lagrangian 4-form 
$\boldsymbol{L}$ into the above definition, we write  the Noether current 3-form $\boldsymbol{J}_{\xi}$ as
\begin{eqnarray}\label{J-38}
&&\ \ \ J_{\alpha_2\alpha_3\alpha_4}\nonumber\\
&&=\epsilon_{\mu\alpha_2\alpha_3\alpha_4}\bigg[
\frac{1}{12}\phi^2g^{\mu\nu}g^{\alpha\beta}(\nabla_{\beta}\mathcal{L}_{\xi} g_{\nu\alpha}-\nabla_{\nu}\mathcal{L}_{\xi} g_{\alpha\beta})\nonumber\\
&&\ \ \ +\frac{1}{3}\phi(\nabla^{\nu}\phi)g^{\beta\mu}\mathcal{L}_{\xi} g_{\nu\beta}-\frac{1}{12}\phi(\nabla^{\mu}\phi)g^{\alpha\beta}\mathcal{L}_{\xi} g_{\alpha\beta}
\nonumber\\
&&\ \ \ -\frac{1}{2}\phi g^{\mu\nu}\nabla_{\nu}\mathcal{L}_{\xi}\phi+\frac{1}{2}g^{\mu\nu}(\nabla_{\nu}\phi)\mathcal{L}_{\xi}\phi\nonumber\\
&&\ \ \ 
-2F^{\mu\alpha a}\mathcal{L}_{\xi} A_{\alpha a}
\bigg]-\xi^{\mu}\epsilon_{\mu\alpha_2\alpha_3\alpha_4}\bigg[\frac{1}{12}R\phi^2\nonumber\\
&&\ \ \ -\frac{1}{2}\phi g^{\alpha\beta}\nabla_{\alpha}\nabla_{\beta}\phi
-\frac{1}{2}F^{a\nu\beta}F_{\nu\beta a}
\bigg].
\end{eqnarray}

We now evaluate each term in the right hand side of  eq. (\ref{J-38}) individually. First, we handle the gravitational contribution involving the Lie derivative of the metric.
Direct calculation shows
\begin{eqnarray}\label{J-39}
&&\ \ \ \frac{1}{12}\phi^2\epsilon_{\mu\alpha_2\alpha_3\alpha_4}
g^{\mu\nu}g^{\alpha\beta}(\nabla_{\beta}\mathcal{L}_{\xi} g_{\nu\alpha}-\nabla_{\nu}\mathcal{L}_{\xi} g_{\alpha\beta})\nonumber\\
&&=\frac{1}{6}\phi^2\epsilon_{\mu\alpha_2\alpha_3\alpha_4}R^{\mu}_{\ \rho}\xi^{\rho}
+\nabla_{\beta}\bigg[\frac{1}{6}\epsilon_{\mu\alpha_2\alpha_3\alpha_4}\phi^2\nabla^{[\beta}\xi^{\mu]}\bigg]
\nonumber\\
&&\ \ \ -\frac{1}{3}\phi(\nabla_{\beta}\phi)\epsilon_{\mu\alpha_2\alpha_3\alpha_4}\nabla^{[\beta}\xi^{\mu]}\nonumber\\
&&=\d \boldsymbol{Q}^g+\frac{1}{6}\phi^2\epsilon_{\mu\alpha_2\alpha_3\alpha_4}R^{\mu}_{\ \rho}\xi^{\rho}\nonumber\\
&&\ \ \ -\frac{1}{3}\phi(\nabla_{\beta}\phi)\epsilon_{\mu\alpha_2\alpha_3\alpha_4}\nabla^{[\beta}\xi^{\mu]},
\end{eqnarray}
where
\begin{eqnarray}
\boldsymbol{Q}^g_{\alpha_3\alpha_4}=-\frac{1}{12}\phi^2\epsilon_{\alpha_3\alpha_4\alpha\mu}\nabla^{[\alpha}\xi^{\mu]}.
\end{eqnarray}

Next we compute the terms associated with the conformal scalar field and metric Lie derivative.
\begin{eqnarray}\label{J-40}
 &&\ \ \epsilon_{\mu\alpha_2\alpha_3\alpha_4}\bigg[\frac{1}{3}\phi(\nabla^{\nu}\phi)g^{\beta\mu}\mathcal{L}_{\xi} g_{\nu\beta}-\frac{1}{12}\phi(\nabla^{\mu}\phi)g^{\alpha\beta}\mathcal{L}_{\xi} g_{\alpha\beta}\bigg]\nonumber\\
&&=\epsilon_{\mu\alpha_2\alpha_3\alpha_4}\bigg[\frac{1}{3}\phi(\nabla^{\nu}\phi)g^{\beta\mu}(\nabla_{\nu}\xi_{\beta}+\nabla_{\beta}\xi_{\nu})\nonumber\\
&&\ \ \  -\frac{1}{6}\phi(\nabla^{\mu}\phi)g^{\alpha\beta}\nabla_{\alpha}\xi_{\beta}\bigg]\nonumber\\
&&=\frac{2}{3}\epsilon_{\mu\alpha_2\alpha_3\alpha_4}\phi(\nabla_{\beta}\phi)\nabla^{(\beta}\xi^{\mu)}\nonumber\\
&&\ \ \ -\frac{1}{6}\epsilon_{\mu\alpha_2\alpha_3\alpha_4}\phi(\nabla^{\mu}\phi)\nabla_{\beta}\xi^{\beta}.
\end{eqnarray}

We then simplify the terms arising from the Lie derivative of the scalar field.
\begin{eqnarray}\label{J-41}
&&\ \ \ \epsilon_{\mu\alpha_2\alpha_3\alpha_4}\bigg[-\frac{1}{2}\phi g^{\mu\nu}\nabla_{\nu}\mathcal{L}_{\xi}\phi+\frac{1}{2}g^{\mu\nu}(\nabla_{\nu}\phi)\mathcal{L}_{\xi}\phi\bigg]\nonumber\\
&&=\epsilon_{\mu\alpha_2\alpha_3\alpha_4}\bigg[-\frac{1}{2}\phi g^{\mu\nu}\nabla_{\nu}(\xi^{\beta}\nabla_{\beta}\phi)\nonumber\\
&&\ \ \ +\frac{1}{2}g^{\mu\nu}(\nabla_{\nu}\phi)(\xi^{\beta}\nabla_{\beta}\phi)\bigg]\nonumber\\
&&=-\frac{1}{2}\phi\epsilon_{\mu\alpha_2\alpha_3\alpha_4}(\nabla^{\mu}\xi^{\beta})(\nabla_{\beta}\phi)
-\frac{1}{2}\phi \epsilon_{\mu\alpha_2\alpha_3\alpha_4}\xi^{\beta}\nabla^{\mu}\nabla_{\beta}\phi\nonumber\\
&&\ \ \ +\frac{1}{2}\epsilon_{\mu\alpha_2\alpha_3\alpha_4}(\nabla^{\mu}\phi)(\xi^{\beta}\nabla_{\beta}\phi).
\end{eqnarray}

For the Yang–Mills gauge sector, we expand the Lie derivative of the gauge potential and rearrange terms to extract another total exterior derivative corresponding to the gauge-field part of the Noether charge as
\begin{eqnarray}\label{J-42}
&&\ \ \ -2\epsilon_{\mu\alpha_2\alpha_3\alpha_4}F^{\mu\alpha a}\mathcal{L}_{\xi}A_{\alpha a}\nonumber\\
&&=-2\epsilon_{\mu\alpha_2\alpha_3\alpha_4}F^{\mu\alpha a}(\xi^{\beta}\nabla_{\beta}A_{\alpha a}+A^{\beta}_a\nabla_{\alpha}\xi_{\beta})\nonumber\\
&&=-4\epsilon_{\mu\alpha_2\alpha_3\alpha_4}F^{\mu\alpha a}\xi^{\beta}\nabla_{[\beta}A_{\alpha]a}\nonumber\\
&&\ \ \ -2\epsilon_{\mu\alpha_2\alpha_3\alpha_4}F^{\mu\alpha a}
\nabla_{\alpha}(A^{\beta}_a\xi_{\beta})\nonumber\\
&&=-4\epsilon_{\mu\alpha_2\alpha_3\alpha_4}F^{\mu\alpha a}\xi^{\beta}\nabla_{[\beta}A_{\alpha]a}\nonumber\\
&&\ \ \ +2\nabla_{\alpha}\big(\epsilon_{\mu\alpha_2\alpha_3\alpha_4}F^{\alpha \mu a}A_a^{\beta}\xi_{\beta}\big)\nonumber\\
&&\ \ \ +2\epsilon_{\mu\alpha_2\alpha_3\alpha_4}A^{\beta}_a\xi_{\beta}\nabla_{\alpha}F^{\mu\alpha a}\nonumber\\
&&=\d\boldsymbol{Q}^{\rm{YM}}
-4\epsilon_{\mu\alpha_2\alpha_3\alpha_4}F^{\mu\alpha a}\xi^{\beta}\nabla_{[\beta}A_{\alpha]a}\nonumber\\
&&\ \ \ +2\epsilon_{\mu\alpha_2\alpha_3\alpha_4}A^{\beta}_a\xi_{\beta}\nabla_{\alpha}F^{\mu\alpha a},
\end{eqnarray}
where
\begin{eqnarray}
\boldsymbol{Q}^{\rm{YM}}_{\alpha_3\alpha_4}=-\epsilon_{\alpha_3\alpha_4\alpha\mu}F^{\alpha\mu a}A^{\beta}_a\xi_{\beta}.
\end{eqnarray}

By substituting all the above intermediate results (\ref{J-39})-(\ref{J-42})  back into (\ref{J-38}), we can obtain the expanded form of the Noether current 3-form
\begin{eqnarray}
&&\ \ J_{\alpha_2\alpha_3\alpha_4}\nonumber\\
&&=(\d Q^g)_{\alpha_2\alpha_3\alpha_4}+(\d Q^{\rm{YM}})_{\alpha_2\alpha_3\alpha_4}\nonumber\\
&&\ \ +\epsilon_{\mu\alpha_2\alpha_3\alpha_4}\bigg[
\frac{1}{6}\phi^2R^{\mu\rho}\xi_{\rho}
-\frac{1}{6}\phi(\nabla_{\beta}\phi)\nabla^{\beta}\xi^{\mu}\nonumber\\
&&\ \ +\frac{1}{6}\phi(\nabla_{\beta}\phi)\nabla^{\mu}\xi^{\beta}
+\frac{1}{3}\phi(\nabla_{\beta}\phi)\nabla^{\beta}\xi^{\mu}\nonumber\\
&&\ \ +\frac{1}{3}\phi(\nabla_{\beta}\phi)\nabla^{\mu}\xi^{\beta}
-\frac{1}{6}\phi(\nabla^{\mu}\phi)\nabla_{\beta}\xi^{\beta}\nonumber\\
&&\ \ -\frac{1}{2}\phi(\nabla^{\mu}\xi^{\beta})\nabla_{\beta}\phi
-\frac{1}{2}\phi\xi^{\beta}\nabla^{\mu}\nabla_{\beta}\phi\nonumber\\
&&\ \ +\frac{1}{2}(\nabla^{\mu}\phi)
(\xi^{\beta}\nabla_{\beta}\phi)
-2F^{\mu\alpha}_a\xi^{\beta}\nabla_{\beta}A^a_{\alpha}\nonumber\\
&&\ \ +2F_a^{\mu\alpha}\xi^{\beta}\nabla_{\alpha}A^a_{\beta}
+2A_a^{\beta}\xi_{\beta}\nabla_{\alpha}F^{\mu\alpha a}\nonumber\\
&&\ \ -\frac{1}{12}\xi^{\mu}R\phi^2+\frac{1}{2}\phi\xi^{\mu}\nabla^{\beta}\nabla_{\beta}\phi
+\frac{1}{2}\xi^{\mu}F^{\alpha\beta a}F_{\alpha\beta a}
\bigg].\ \ 
\end{eqnarray}

We further simplify this expression by making use of the field equations of motion (\ref{E-28})-(\ref{E-30}) derived in the preceding section. 
After algebraic rearrangement and extraction of an additional total divergence term contributed by the scalar field, the Noether current 3-form $J_{\alpha_2\alpha_3\alpha_4}$ is finally decomposed into
\begin{eqnarray}
&&\ \ \ J_{\alpha_2\alpha_3\alpha_4}\nonumber\\
&&=(\d \bQ^g)_{\alpha_2\alpha_3\alpha_4}+(\d \bQ^{\rm{YM}})_{\alpha_2\alpha_3\alpha_4}\nonumber\\
&&\ \ \ +\epsilon_{\mu\alpha_2\alpha_3\alpha_4}\bigg[
2E^{\mu\rho}_g\xi_{\rho}-E^{\mu}_{\rm{YM}}A^{\beta}_a\xi_{\beta}\nonumber\\
&&\ \ \ +\nabla_{\beta}\big(\frac{1}{3}\phi(\nabla^{[\beta}\phi)\xi^{\mu]}\big)\bigg]\nonumber\\
&&=(\boldsymbol{C}_{\xi})_{\alpha_2\alpha_3\alpha_4}+(\d \bQ_{\xi})_{\alpha_2\alpha_3\alpha_4}.
\end{eqnarray}
In the above decomposition, $\boldsymbol{C}_{\xi}$
 is referred to as the constraint 3-form, and its explicit component reads
\begin{eqnarray}
&&(\boldsymbol{C}_{\xi})_{\alpha_2\alpha_3\alpha_4}=\bep_{\mu\alpha_2\alpha_3\alpha_4}(2E^{\mu\beta}_g\xi_{\beta}-E^{\mu}_{\rm{YM}}A_a^{\beta}\xi_{\beta}).
\end{eqnarray}
The quantity $\bQ_{\xi}$ is the  Noether charge 2-form
\begin{eqnarray}
&&\bQ_{\xi}=\bQ_{\xi}^g+\bQ_{\xi}^{\phi}+\bQ_{\xi}^{\rm{YM}},
\end{eqnarray}
where
\begin{eqnarray}
&&\boldsymbol{Q}^g_{\alpha_3\alpha_4}=-\frac{1}{12}\phi^2\epsilon_{\alpha_3\alpha_4\alpha\mu}\nabla^{\alpha}\xi^{\mu},\label{Q49}\\
&&\boldsymbol{Q}^{\phi}_{\alpha_3\alpha_4}=-\frac{1}{6}\phi \epsilon_{\alpha_3\alpha_4\alpha\mu}(\nabla^{\alpha}\phi)\xi^{\mu},\label{Q50}\\
&&\boldsymbol{Q}^{\rm{YM}}_{\alpha_3\alpha_4}=-\epsilon_{\alpha_3\alpha_4\alpha\mu}F^{\alpha\mu a}A^{\beta}_a\xi_{\beta}.\label{Q51}
\end{eqnarray}

When evaluated on the on-shell solutions, one has $\boldsymbol{C}_{\xi}=0$ and
\begin{eqnarray}
J_{\alpha_2\alpha_3\alpha_4}=(\d\boldsymbol{Q}_{\xi})_{\alpha_2\alpha_3\alpha_4}.
\end{eqnarray}

\section{First Law of black hole mechanics in CEPYM theory}\label{sec3}

Based on the symplectic potential 3-form and the Noether charge 2-form, we derive the first law of black hole mechanics for the CEPYM theory in this section.

Let $\xi^{\mu}$ be an arbitrary fixed vector field on $M$. We then have
\begin{equation}
\begin{split}  
\delta\boldsymbol{J}_{\xi}&=\delta\boldsymbol{\Theta}(\tilde{\phi},\mathcal{L}_{\xi}\tilde{\phi})-\xi\cdot\delta\boldsymbol{L}\\
&=\delta \boldsymbol{\Theta}(\tilde{\phi},\mathcal{L}_{\xi}\tilde{\phi})-\xi\cdot \d\boldsymbol{\Theta}(\tilde{\phi},\delta\tilde{\phi})-\xi\cdot \boldsymbol{E}^{\tilde{\phi}}\delta\tilde{\phi}\\
&=\delta \boldsymbol{\Theta}(\tilde{\phi},\mathcal{L}_{\xi}\tilde{\phi})-\mathcal{L}_{\xi}\boldsymbol{\Theta}(\tilde{\phi},\delta\tilde{\phi})\nonumber\\
&\ \ \ \ +\d\big(\xi\cdot\boldsymbol{\Theta}(\tilde{\phi},\delta\tilde{\phi})\big)
-\xi\cdot \boldsymbol{E}^{\tilde{\phi}}\delta\tilde{\phi}.
\end{split}  
\end{equation}
Here $\tilde{\phi}=(g_{\mu\nu},\phi,A^a_{\mu})$ denotes the full set of dynamical fields. In the above derivation,  we have used the relations
\begin{eqnarray}
\delta \boldsymbol{L}=\d\boldsymbol{\Theta}+\boldsymbol{E}^{\tilde{\phi}}\delta\tilde{\phi}
\end{eqnarray}
and
\begin{eqnarray}
\mathcal{L}_{\xi}\boldsymbol{\Theta}=\xi\cdot\d\boldsymbol{\Theta}+\d(\xi\cdot \boldsymbol{\Theta}).
\end{eqnarray}

Following the definition of the symplectic 3-form $\boldsymbol{\omega}$
 given in \cite{IW1994}, we have
\begin{eqnarray}
\boldsymbol{\omega}(\tilde{\phi},\delta\tilde{\phi},\mathcal{L}_{\xi}\tilde{\phi})=\delta\boldsymbol{\Theta}(\tilde{\phi},\mathcal{L}_{\xi}\tilde{\phi})-\mathcal{L}_{\xi}\boldsymbol{\Theta}(\tilde{\phi},\delta\tilde{\phi}),
\end{eqnarray}
which yields
\begin{equation}\label{ther1}
\begin{split}
    &\ \ \ \boldsymbol{\omega}(\tilde{\phi},\delta\tilde{\phi},\mathcal{L}_{\xi}\tilde{\phi})\\
    &=\delta\boldsymbol{J}_{\xi}-\d(\xi\cdot \boldsymbol{\Theta})-\xi\cdot \boldsymbol{E}^{\tilde{\phi}}\delta\tilde{\phi}\\
    &=\delta\boldsymbol{C}_{\xi}+\delta\d\bQ_{\xi}-\d(\xi\cdot \boldsymbol{\Theta})-\xi\cdot \boldsymbol{E}^{\tilde{\phi}}\delta\tilde{\phi}.
    \end{split}
\end{equation}

If $\xi^{\mu}$  is a Killing vector field that generates a symmetry of all dynamical fields, 
and $\delta\tilde{\phi}$ satisfies the linearized equations of motion, Eq.(\ref{ther1}) becomes 
\begin{eqnarray}\label{ther2}
\d\delta\boldsymbol{Q}_{\xi}-\d(\xi\cdot\boldsymbol{\theta})=0.
\end{eqnarray}

Let $\Sigma$ be a spacelike hypersurface which extends to infinity and has an inner boundary $\partial\Sigma$ in an asymptotically flat spacetime. Integrating eq.(\ref{ther2}) over the hypersurface $\Sigma$, we obtain
\begin{eqnarray}
\int_{\partial\Sigma}\delta\boldsymbol{Q}_{\xi}-\xi\cdot\boldsymbol{\Theta}(\tilde{\phi},\delta\tilde{\phi})=\int_{\infty}\delta\boldsymbol{Q}_{\xi}-\xi\cdot\boldsymbol{\Theta}(\tilde{\phi},\delta\tilde{\phi}).
\end{eqnarray}

Let $(g_{\mu\nu},A^a_{\mu},\phi)$ be a stationary axisymmetric black hole solution to the equations of motion derived from the conformal Einstein-power-Yang-Mills Lagrangian. If the  black hole spacetime possesses a bifurcation surface,  we denote the horizon Killing field of the stationary axisymmetric black hole  by
$$\xi^{\mu}=t^{\mu}+\Omega_H\varphi^{\mu}.$$

Let $\Sigma$ be an hypersurface connecting infinity and a cross section on the horizon. Denote the inner boundary of $\Sigma$ by $S_{H}$. Then we have
\begin{eqnarray}\label{1stlaw-0}
\int_{S_H}\delta\boldsymbol{Q}_{\xi}-\xi \cdot\boldsymbol{\Theta}(\tilde{\phi},\delta\tilde{\phi})=\int_{\infty}\delta\boldsymbol{Q}_{\xi}-\xi\cdot\boldsymbol{\Theta}(\tilde{\phi},\delta\tilde{\phi}).
\end{eqnarray}

Consider a perturbation $\delta\tilde{\phi}$ which generates a slightly different stationary black hole.  When comparing two nearby spacetimes, there is a certain freedom in which points are chosen to correspond. Here we adopt the gauge choice as \cite{Jiangjie2021}
\begin{eqnarray}
    \delta\xi^{\mu}=0.
\end{eqnarray}

\subsection{Asymptotic evaluation at spatial infinity}

We first evaluate the boundary integral terms at spatial infinity.
\begin{equation}
    \begin{split}
&\int_{\infty}\delta\boldsymbol{Q}_{\xi}-\xi\cdot\boldsymbol{\Theta} \\
=&\int_{\infty}(\delta\boldsymbol{Q}_{t}-t\cdot\boldsymbol{\Theta})
+\int_{\infty}(\Omega_H\delta\boldsymbol{Q}_{\varphi}-\Omega_H\varphi\cdot\boldsymbol{\Theta})\\
=&\int_{\infty}(\delta\boldsymbol{Q}_{t}-t\cdot\boldsymbol{\Theta})
+\int_{\infty}\Omega_H\delta\boldsymbol{Q}_{\varphi}\\
=&\delta\mathcal{E}-\Omega_H\delta \mathcal{J},
    \end{split}
\end{equation}
where $\mathcal{E}$ and $\mathcal{J}$ denote the canonical energy and canonical angular momentum of the asymptotically flat spacetime, respectively \cite{Gao2003}. More precisely, we have
\begin{equation}
\begin{split}
\delta\mathcal{E}&=\int_{\infty}(\delta \boldsymbol{Q}_t-t\cdot\boldsymbol{\Theta})\\
&=\int_{\infty}(\delta \boldsymbol{Q}_t^g+\delta \boldsymbol{Q}_t^{\phi}
+\delta \boldsymbol{Q}_t^{\rm{YM}})\\
&\ \ \ -\int_{\infty}t\cdot(\boldsymbol{\Theta}^g+
\boldsymbol{\Theta}^{\phi}+\boldsymbol{\Theta}^{\rm{YM}}),\\
\end{split}
\end{equation}
and
\begin{eqnarray}
    \delta\mathcal{J}&=-\int_{\infty}\delta \boldsymbol{Q}_{\varphi}
    =-\int_{\infty} \delta (\boldsymbol{Q}_{\varphi}^g+\boldsymbol{Q}_{\varphi}^{\phi}+\boldsymbol{Q}_{\varphi}^{\rm{YM}}).
\end{eqnarray}

\subsubsection{Calculation of  $\delta\mathcal{E}$}

Consider spacetimes which are asymptotically ﬂat in the sense that there exists a ﬂat metric $eta_{\mu\nu}$  such that in a 
global inertial coordinate system $\{t,x^1,x^2,x^3\}$ of $\eta_{\mu\nu}$ we have
\begin{eqnarray}
&&\phi=1+\frac{\phi_1}{r}+\frac{\phi_2}{r^2}+\cdots,\label{phi-64}\\
&&g_{\mu\nu}=\eta_{\mu\nu}+h_{\mu\nu}, h_{\mu\nu}=O(\frac{1}{r}),\partial_\alpha h_{\mu\nu}=O(\frac{1}{r^2}), 
\end{eqnarray}
where $r=\sqrt{\sum\limits_{i=1}^3(x^i)^2.}$
Direct calculation shows
\begin{equation}
\begin{split}
&\ \ \ \lim_{r\rightarrow\infty}\int_{S_r}-\frac{1}{12}\phi^2\epsilon_{\alpha_3\alpha_4\alpha\mu}\nabla^{\alpha}t^{\mu}\\
&=-\frac{1}{12}\lim_{r\rightarrow\infty}\int_{S_r}\epsilon_{\alpha_3\alpha_4\alpha\mu}g^{\alpha\beta}\nabla_{\beta}(\frac{\partial}{\partial t})^{\mu}\\
&=-\frac{1}{12}\lim_{r\rightarrow\infty}\int_{S_r}\epsilon_{\alpha_3\alpha_4\alpha\mu}g^{\alpha\beta}\Gamma^{\mu}_{\beta\rho}(\frac{\partial}{\partial t})^{\rho}\\
&=-\frac{1}{12}\lim_{r\rightarrow\infty}\int_{S_r}\epsilon_{\alpha_3\alpha_4\alpha\mu}g^{\alpha\beta}\Gamma^{\mu}_{\beta t}\\
&=-\frac{1}{12}\lim_{r\rightarrow\infty}\int\bigg(\frac{\partial g_{tt}}{\partial r}-\frac{\partial g_{rt}}{\partial t}\bigg)\d S,
\end{split}
\end{equation}

and

\begin{equation}
\begin{split}
&\lim_{r\rightarrow\infty}\int_{S_r}\frac{\phi^2}{12}t^{\alpha_2}\epsilon_{\mu\alpha_2\alpha_3\alpha_4}
 g^{\mu\nu}g^{\alpha\beta}(\nabla_{\beta}\delta g_{\nu\alpha}-\nabla_{\nu}
 \delta g_{\alpha\beta})\\
 &=-\lim_{r\rightarrow\infty}\int_{S_r}\frac{\phi^2}{12}\epsilon_{t\mu\alpha_3\alpha_4}
 g^{\mu\nu}g^{\alpha\beta}(\nabla_{\beta}\delta g_{\nu\alpha}-\nabla_{\nu}\delta g_{\alpha\beta})\\
 &=-\lim_{r\rightarrow\infty}\int_{S_r}\frac{\phi^2}{12}(\d r)_{\mu}g^{\mu\nu}
 g^{\alpha\beta}(\nabla_{\beta}\delta g_{\nu\alpha}-\nabla_{\nu}\delta g_{\alpha\beta})\d S\\
 &=-\lim_{r\rightarrow\infty}\int_{S_r}\frac{\phi^2}{12}r^{\nu}
 \big[-(\partial_t\delta g_{\nu t}-\partial_{\nu}\delta g_{tt})\\
 &\ \ \ \ \ \ \ \ \ \ \ \ \ \ \ \ \ \ \ \ +(\partial_j\delta g_{\nu j}-\partial_{\nu}g_{jj})\big]\d S\\
 &=-\lim_{r\rightarrow\infty}\int_{S_r}\frac{\phi^2}{12}
 \big[\partial_{r}\delta g_{tt}-(\partial_t\delta g_{r t})\\
 &\ \ \ \ \ \ \ \ \ \ \ \ \ \ \ \ \ \ \ \ +r^i(\partial_j\delta g_{i j}-\partial_{i}g_{jj})\big]\d S\\
 &=\delta\bigg[-\frac{1}{12}\lim_{r\rightarrow\infty}\int_{S_r}
 \big[\partial_{r}\delta g_{tt}-(\partial_t\delta g_{r t})\\
 &\ \ \ \ \ \ \ \ \ \ \ \ \ \ \ \ \ \ \ \ +r^i(\partial_j\delta h_{i j}-\partial_{i}h_{jj})\big]\d S\bigg],\nonumber
 \end{split}
\end{equation}
where $r^{\nu}=(\frac{\partial}{\partial r})^{\nu},\ \d S=\sqrt{-g}\d\theta\wedge\d\varphi.$

Hence, we have
\begin{equation}
    \begin{split}
 &\ \ \ \int_{\infty}\delta\boldsymbol{Q}_{t}^g-t\cdot\boldsymbol{\Theta}^g  
 =\delta \mathcal{M},
    \end{split}
\end{equation}
where \begin{eqnarray}
    \mathcal{M}=\frac{1}{12}\int_{\infty}(\partial_ih_{jj}-\partial_jh_{ij})r^i
\end{eqnarray}
is the ADM mass of the spacetime.

We next calculate $\int_{\infty}\delta\boldsymbol{Q}_{t}^{\phi}-t\cdot\boldsymbol{\Theta}^{\phi}.$ Define the scalar charge $q$ associated with the conformal scalar field $\phi$ as
\begin{equation}
\begin{split}
q:=&\int_{\infty}\boldsymbol{Q}_t^{\phi}\\
=&-\frac{1}{6}\int_{\infty}\phi\epsilon_{\alpha_3\alpha_4\alpha\mu}(\nabla^{\alpha})t^{\mu}\\
=&-\frac{1}{6}\int_{\infty}\epsilon_{\alpha t\alpha_3\alpha_4}\nabla^{\alpha}\phi\\
=&-\frac{1}{6}\int_{S^2}\phi_1\sin\theta\d\theta\d\varphi.
\end{split}
\end{equation}
For $\phi=1+\phi_2/r^2+\cdots$, one has $q=0$. Furthermore,
\begin{equation}
\begin{split}
    &\ \ \ \int_{\infty}-t\cdot\boldsymbol{\Theta}^{\phi}\\
   &= -\int_{\infty}t^{\alpha_2}\epsilon_{\mu\alpha_2\alpha_3\alpha_4}\bigg(\frac{1}{3}
   \phi(\nabla^{\nu}\phi)g^{\beta\mu}\delta g_{\nu\beta}
   \\
   &\ \ \ \  -\frac{1}{12}\phi(\nabla^{\mu}\phi)g^{\alpha\beta}\delta g_{\alpha\beta}-\frac{1}{2}\phi g^{\mu\nu}\nabla_{\nu}\delta\phi\\
   &\ \ \ \ +
   \frac{1}{2}g^{\mu\nu}(\nabla_{\nu}\phi)\delta\phi\bigg)\\
   &=\frac{1}{2}\int_{S^2}(\delta\phi_1)\sin\theta\d\theta\d\varphi=-3\delta q.
\end{split}
\end{equation}
So we have
\begin{eqnarray}
\int_{\infty}\delta\boldsymbol{Q}_{t}^{\phi}-t\cdot\boldsymbol{\Theta}^{\phi}=-2\delta q.
\end{eqnarray}

To calculate $\int_{\infty}\delta\boldsymbol{Q}_{t}^{\rm{YM}}-t\cdot\boldsymbol{\Theta}^{\rm{YM}},$ we
define the asymptotic Yang–Mills potential magnitude $V$ and Yang–Mills charge $Q^\infty$ following \cite{Gao2003}:
\begin{eqnarray}
V:=\lim_{r\rightarrow\infty}(A^a_tA_ta)^{1/2},
E^{\mu}_a:=\sqrt{h}F^{\mu\nu}_an_{\nu},
\end{eqnarray}
where $n^{\nu}$ is the unit normal to the spacelike hypersurface $\Sigma$. One has
\begin{equation}
\begin{split}
\int_{\infty}\boldsymbol{Q}_t^{\rm{YM}}&=-\int_{\infty}\epsilon_{\alpha_3\alpha_4\alpha\mu}
F^{\alpha\mu}_aA^a_{\beta}t^{\beta}\\
&=-\int_{\infty}\epsilon_{\alpha_3\alpha_4\alpha\mu}F^{\alpha\mu}_aA^a_t.
\end{split}
\end{equation}
As shown in \cite{Gao2003},
\begin{eqnarray}
\int_{\infty}\boldsymbol{Q}_t^{\rm{YM}}=VQ^{\infty},
\end{eqnarray}
where
\begin{eqnarray}
Q^{\infty}:=\frac{1}{2}\int_{\infty}|E^{\mu}_ar_{\mu}|
\end{eqnarray}
is the Yang–Mills charge, with $r^\mu$ the unit radial vector and $|\cdot|$ the Lie algebra norm. In addition,
\begin{equation}
\begin{split}
\int_{\infty}t\cdot\boldsymbol{\Theta}
^{\rm{YM}}&=-2\int_{\infty}\epsilon_{\mu\alpha_2\alpha_3\alpha_4}t^{\alpha_2}F^{\mu\alpha}_a\delta A^a_{\alpha}\\
 &=Q^{\infty}\delta V.
\end{split}
\end{equation}
This yields
\begin{eqnarray}
\int_{\infty}\delta\boldsymbol{Q}^{\rm{YM}}-t\cdot\boldsymbol{\Theta}^{\rm{YM}}=V\delta Q^{\infty}.
\end{eqnarray}
Combining all asymptotic contributions, the variation of the canonical energy is summarized as
\begin{eqnarray}\label{deltaE}
\delta\mathcal{E}=\delta\mathcal{M}-2\delta q+V\delta Q^{\infty}.
\end{eqnarray}

For the trivial background scalar field $\phi=1$, the scalar charge vanishes, and the result reduces to $$\delta\mathcal{E}=\delta\mathcal{M}+V\delta Q^{\infty},$$ which perfectly recovers the classic results in \cite{Gao2003,HS1993}.

\subsubsection{Calculation of the canonical angular momentum}

 By using eqs. (\ref{Q49}) and (\ref{phi-64}), one can obtain
\begin{equation}
\begin{split}
    -\int_{\infty}\boldsymbol{Q}_{\varphi}^g&=\int_{\infty}\frac{1}{12}\phi^2\epsilon_{\alpha_3\alpha_4\alpha\mu}\nabla^{\alpha}\varphi^{\mu}\\
    &
    =\frac{1}{12}\int_{\infty}\epsilon_{\alpha_3\alpha_4\alpha\mu}\nabla^{\alpha}\varphi^{\mu},
\end{split}
\end{equation}
which is just the expression for the angular momentum in the vacuum asymptotically flat spacetime.

In addition, it follows from eqs.(\ref{Q50}) and (\ref{phi-64}) that
\begin{equation}
\begin{split}
&-\int_{\infty}\boldsymbol{Q}_{\varphi}^{\phi}=\frac{1}{6}\int_{\infty}\phi\epsilon_{\alpha_3\alpha_4\alpha\mu}(\nabla^{\alpha}\phi)\varphi^{\mu}=0.
\end{split}
\end{equation}

Furthermore, eq. (\ref{Q51}) yields
\begin{eqnarray}
-\int_{\infty}\delta\boldsymbol{Q}_{\varphi}^{\rm{YM}}=\int_{\infty}\epsilon_{\alpha_3\alpha_4\alpha\mu}
F^{\alpha\mu}_aA^a_{\beta}\xi^{\beta}.
\end{eqnarray}

The total canonical angular momentum thus reads
\begin{eqnarray}\label{J-80}
\mathcal{J}=\frac{1}{12}\int_{\infty}\epsilon_{\alpha_3\alpha_4\alpha\mu}\nabla^{\alpha}\varphi^{\mu}+\int_{\infty}\epsilon_{\alpha_3\alpha_4\alpha\mu}
F^{\alpha\mu}_a A^a_{\beta}\xi^{\beta}.
\end{eqnarray}
This expression implies that the gravitational and Yang–Mills sectors contribute to the total canonical angular momentum, while the scalar field contribution vanishes asymptotically. Eq. (\ref{J-80}) is consistent with the canonical angular momentum formula for axisymmetric black holes in Einstein-Yang-Mills theory \cite{SW1992,Gao2003}.

\subsection{Near-horizon evaluation}

We now evaluate the horizon boundary integral in eq.(\ref{1stlaw-0}) by adopting null Gaussian normal coordinates adapted to the black hole horizon \cite{HIW2007}. The near-horizon metric and its inverse take explicit diagonal-block forms, and the horizon Killing vector $\xi^\mu$ is null on $\mathcal{H}$ with surface gravity $\kappa$.

Within the null Gaussian normal coordinates $\{v,r,\theta^A\} (\theta^A=\theta,\varphi)$, the line element 
near the horizon $\mathcal{H}$ reads \cite{HIW2007}
\begin{eqnarray}
\d s^2=2(\d r-r\alpha\d v-r\beta_A\d\theta^A)\d v+\gamma_{AB}\d\theta^A\d\theta^B,
\end{eqnarray}
where $\alpha,\beta_A$ and $\gamma_{AB}$  are functions of $r$ and $\theta^A$. On the horizon $\mathcal{H}$, $r=0$ and $\xi^{\mu}=(\frac{\partial}{\partial v})^{\mu}$ is  null vector there. In other words,
\begin{eqnarray}\label{2.80}
&&g_{\mu\nu}=\begin{pmatrix}
 -2r\alpha & 1 & -r\beta_{\theta} & -r\beta_{\varphi} \\
  1& 0 & 0 & 0\\
  -r\beta_{\theta}& 0 &\gamma_{\theta\theta}  &\gamma_{\theta\varphi} \\
  -r\beta_{\varphi}&0  &\gamma_{\theta\varphi}  &\gamma_{\varphi\varphi}
\end{pmatrix},\\
&&g^{\mu\nu}=\begin{pmatrix}
 0 & 1 & 0 & 0\\
1& g^{rr} & g^{r\theta}  & r^{r\varphi}\\
 0 & g^{r\theta} & \gamma^{\theta\theta} &\gamma^{\theta\varphi} \\
  0& g^{r\varphi} &\gamma^{\theta\varphi}  &\gamma^{\varphi\varphi}
\end{pmatrix},
\end{eqnarray}
where $\gamma^{AB}$ is the inverse matrix of $\gamma_{AB}.$ On the horizon $\mathcal{H}$,
\begin{equation}\label{2.82}
\begin{split}
&g_{\mu\nu}|_{\mathcal{H}}=\begin{pmatrix}
 0 & 1 & 0 & 0 \\
  1& 0 & 0 & 0\\
  0& 0 &\gamma_{\theta\theta}  &\gamma_{\theta\varphi} \\
  0&0  &\gamma_{\theta\varphi}  &\gamma_{\varphi\varphi}
\end{pmatrix},\\
&g^{\mu\nu}|_{\mathcal{H}}=\begin{pmatrix}
 0 & 1 & 0 & 0\\
1& 0 & 0 & 0\\
 0 & 0 & \gamma^{\theta\theta} &\gamma^{\theta\varphi} \\
  0& 0&\gamma^{\theta\varphi}  &\gamma^{\varphi\varphi}
\end{pmatrix}.
\end{split}
\end{equation}
Let
\begin{eqnarray}
    s^{\mu}=(\frac{\partial}{\partial r})^a,
\end{eqnarray}
then $s^{\mu}$ is another null vector field on $\mathcal{H}$. Moreover, on $\mathcal{H}$, we have
\begin{eqnarray}
\nabla_{\mu}\xi_{\nu}=\kappa\hat{\epsilon}_{\mu\nu}-\beta_A\xi_{[\mu}(\d\theta^A)_{\nu]},
\end{eqnarray}
where $\kappa=\alpha|_{H}$ is the surface gravity of $\mathcal{H}$ and 
\begin{eqnarray}
\hat{\epsilon}_{\mu\nu}=(\d v\wedge \d r)_{\mu\nu}=(\d v)_{\mu}(\d r)_{\nu}
-(\d r)_{\mu}(\d v)_{\nu}
\end{eqnarray}
is the binormal to $\mathcal{H}$.
Then
\begin{eqnarray}
&&\ \ \int_{S_H}Q^g_{\alpha_3\alpha_4}[\xi]\nonumber\\
&&=\int_{S_H}-\frac{1}{12}\phi^2\epsilon_{\alpha_3\alpha_4\alpha\mu}\nabla^{\alpha}\xi^{\mu}\nonumber\\
&&=-\frac{1}{12}\int_{S_H}\phi^2\epsilon_{\alpha_3\alpha_4\alpha\mu}
(\kappa\hat{\epsilon}^{\alpha\mu}-\beta_A\xi^{\alpha}(\d\theta^A)^{\mu})\nonumber\\
&&=-\frac{\kappa\phi_H^2}{12}\int_{S_H}\epsilon_{\alpha_3\alpha_4\alpha\mu}\hat{\epsilon}^{\alpha\mu}\nonumber\\
&&=\frac{\phi_H^2}{6}\kappa A,
\end{eqnarray}
where $A$ is area of the black hole. Hence 
\begin{equation}
    \begin{split}
{\delta}\int_{S_H}\boldsymbol{Q}_{\xi}^g=\frac{1}{6}\delta(\phi_H^2\kappa A).
    \end{split}
\end{equation}
Furthermore,
\begin{equation}\nonumber
\begin{split}
&\ \ \int_{S_H}\xi\cdot\boldsymbol{\Theta}^g\\
&=\int_{S_H}
\frac{\phi^2}{12}\xi^{\alpha_2}\epsilon_{\mu\alpha_2\alpha_3\alpha_4}g^{\mu\nu}g^{\alpha\beta}
(\nabla_{\beta}\delta g_{\nu\alpha}-\nabla_{\nu}\delta g_{\alpha\beta})\\
&=\frac{1}{12}\int_{S_H}\phi^2\xi^{\alpha_2}\epsilon_{\mu\alpha_2\alpha_3\alpha_4}
(g^{\mu\nu}g^{\alpha\beta}-g^{\mu\beta}g^{\nu\alpha})\nabla_{\beta}\delta g_{\nu\alpha}\\
&=\frac{1}{12}\int_{S_H}\phi^2\xi^{\alpha_2}\epsilon_{\mu\alpha_2\alpha_3\alpha_4}
(g^{\mu\nu}g^{\alpha\beta}-g^{\mu\beta}g^{\nu\alpha})(\partial_{\beta}\delta g_{\nu\alpha}\\
&\ \ \ \ \ \ \ \ -\Gamma^{\rho}_{\beta\nu}\delta g_{\rho\alpha}-
\Gamma^{\rho}_{\beta\alpha}\delta g_{\nu\rho})\\
&=\frac{1}{12}\int_{S_H}\phi^2\epsilon_{rv\alpha_3\alpha_4}\partial_r\delta g_{vv}\\
&=\frac{1}{6}\int_{S_H}\phi^2\epsilon_{vr\alpha_3\alpha_4}\delta\alpha\\
&=\frac{1}{6}\int_{S_H}\phi^2\epsilon_{vr\alpha_3\alpha_4}\delta\kappa\\
&=\frac{1}{6}\phi_H^2A\delta\kappa,
\end{split}
\end{equation}
where we have used equations (\ref{2.80}) and (\ref{2.82}). Therefore,
\begin{eqnarray}
\int_{S_H}\delta\boldsymbol{Q}_{\xi}^g-\xi\cdot\boldsymbol{\Theta}^g
=\frac{1}{6}\phi_H^2\kappa\delta A+\frac{1}{3}\kappa A\phi_H\delta\phi_H.
\end{eqnarray}

We next compute $\int_{S_H}\delta\boldsymbol{Q}_{\xi}^{\phi}-\xi\cdot\boldsymbol{\Theta}^{\phi}.$ Direct calculation shows
\begin{equation}
\begin{split}
 \int_{S_H}\boldsymbol{Q}_{\xi}^{\phi}&=\int_{S_H}-\frac{1}{6}\epsilon_{\alpha_3\alpha_4\alpha\mu}
 (\nabla^\alpha\phi)\xi^{\mu}\\
 &=-\frac{1}{6}\int_{S_H}\phi\epsilon_{\alpha\mu\alpha_3\alpha_4}g^{\alpha\beta}(\nabla_{\beta}\phi)\xi^{\mu}\\
 &=-\frac{1}{6}\int_{S_H}\phi \epsilon_{\alpha v\alpha_3\alpha_4}g^{\alpha\beta}\nabla_{\beta}\phi\\
 &=-\frac{1}{6}\int_{S_H}\phi\epsilon_{rv\alpha_3\alpha_4}g^{r\beta}\nabla_{\beta}\phi\\
 &=\frac{1}{6}\int_{S_H}\phi\epsilon_{vr\alpha_3\alpha_4}\nabla_v\phi\\
 &=0,
\end{split}
\end{equation}
and
\begin{equation}
    \begin{split}
&\ \ \int_{S_H}\xi\cdot\boldsymbol{\Theta}^{\phi}\\
&=\int_{S_H}\xi^{\alpha_2}
\epsilon_{\mu\alpha_2\alpha_3\alpha_4}\bigg(
\frac{1}{3}\phi(\nabla^{\nu}\phi)g^{\beta\mu}\delta g_{\nu\beta}\\
&\ \ \ \ \ \ \ \ \ -\frac{1}{12}\phi(\nabla^{\mu}\phi)g^{\alpha\beta}\delta g_{\alpha\beta}\\
&\ \ \ \ \ \ \ \ \ -\frac{1}{2}g^{\mu\nu}\nabla_{\nu}\delta\phi+\frac{1}{2}
g^{\mu\nu}(\nabla_{\nu}\phi)\delta\phi\bigg)\\
&=\int_{S_H}\xi^{\alpha_2}
\epsilon_{rv\alpha_3\alpha_4}\bigg(
\frac{1}{3}\phi(\nabla^{\nu}\phi)g^{\beta r}\delta g_{\nu\beta}\\
&\ \ \ \ \ \ \ \ \ -\frac{1}{12}\phi g^{\rho r}(\nabla_{\rho}\phi)g^{\alpha\beta}\delta g_{\alpha\beta}\\
&\ \ \ \ \ \ \ \ \ -\frac{1}{2}g^{r\nu}\nabla_{\nu}\delta\phi+\frac{1}{2}
g^{r\nu}(\nabla_{\nu}\phi)\delta\phi\bigg)\\
&=0.
    \end{split}
\end{equation}
Hence,
\begin{eqnarray}
\int_{S_H}\delta\boldsymbol{Q}_{\xi}^{\phi}-\xi\cdot\boldsymbol{\Theta}^{\phi}
=0.
\end{eqnarray}

Finally, we calculate $\int_{S_H}\delta\boldsymbol{Q}_{\xi}^{\rm{YM}}-\xi\cdot\boldsymbol{\Theta}^{\rm{YM}}.$ Direct computation shows
\begin{equation}
\begin{split}
&\ \ \ \int_{S_H}\delta\boldsymbol{Q}_{\xi}^{\rm{YM}}-\xi\cdot\boldsymbol{\Theta}^{\rm{YM}}\\
&=\int_{S_H}\delta\big(-\epsilon_{\alpha_3\alpha_4\alpha\mu}F^{\alpha\mu}_a
A^a_{\beta}\xi^{\beta}\big)\\
&\ \ \ \ \ \ \ \ \ -\xi^{\alpha_2}\big(-2\epsilon_{\mu\alpha_2\alpha_3\alpha_4}F^{\mu\alpha}_a\delta A^a_{\alpha}\big)\\
&=-\int_{S_H}A_{\beta}^a\xi^{\beta}\delta(\epsilon_{\alpha_3\alpha_4\alpha\mu}F^{\alpha\mu}_a)\\
&\ \ \ \ -\int_{S_H}\epsilon_{\alpha_3\alpha_4\alpha\mu}F^{\alpha\mu}_a\xi^{\beta}\delta A_{\beta}^a\\
&\ \ \ \ +\int_{S_H}2\xi^{\alpha_2}\epsilon_{\mu\alpha_2\alpha_3\alpha_4}F^{\mu\alpha}_a\delta A_{\alpha}^a\\
&=-\int_{S_H}A_{\beta}^a\xi^{\beta}\delta(\epsilon_{\alpha_3\alpha_4\alpha\mu}F^{\alpha\mu}_a),
\end{split}
\end{equation}
where we have assumed the condition $F_a^{rB}|_{\mathcal{H}}=0$ which was used in \cite{Gao2003}.

For electromagnetic field, the quantity $\Phi^{\rm{EM}}:=-A_{\mu}\xi^{\mu}|_{\mathcal{H}}$ is a constant 
on the horizon, and
\begin{eqnarray}
\int_{S_H}\epsilon_{\alpha_3\alpha_4\alpha\mu}F^{\alpha\mu} =Q^{\rm{EM}}
\end{eqnarray}
is the electric charge of the black hole spacetime. Then we have
\begin{eqnarray}
\int_{S_H}\delta\boldsymbol{Q}_{\xi}^{\rm{EM}}-\xi\cdot\boldsymbol{\Theta}^{\rm{EM}}=\Phi^{\rm{EM}}\delta Q^{\rm{EM}}.
\end{eqnarray}
For the Yang-Mills field,  in general, the integral $-\int_{S_H}A_{\beta}^a\xi^{\beta}\delta(\epsilon_{\alpha_3\alpha_4\alpha\mu}F^{\alpha\mu}_a)$ can not be implied further to the form $\Phi\delta Q$ as the electromagnetic case. This is due to the complexity of $SU(2)$ Lie algebra \cite{HS1993}.

\subsection{The first law of black hole mechanics}

Collecting the above results obtained in previous subsections, one can get finally that
\begin{eqnarray}\label{1stlaw}
&&\delta\mathcal{E}=\Omega_H\delta\mathcal{J}+
\frac{1}{6}\kappa\delta(\phi_H^2 A)\nonumber\\
&&\ \ \ \ \ \ 
-\int_{S_H}A_{\beta}^a\xi^{\beta}\delta(\epsilon_{\alpha_3\alpha_4\alpha\mu}F^{\alpha\mu}_a).
\end{eqnarray}

Recall that the Wald entropy is defined by  \cite{IW1994}
\begin{eqnarray}
S:=-2\pi\int_{S_H}\epsilon_{\alpha_3\alpha_4}\frac{\delta L}{\delta R_{\mu\nu\alpha\beta}}\hat\epsilon_{\mu\nu}\hat{\epsilon}_{\alpha\beta}.
\end{eqnarray}
For CEPYM theory, the Lagrange density is given by (\ref{lagrangeL}). Direct calculation shows
\begin{eqnarray}
  S  =\frac{\pi}{3}\phi_H^2A.
\end{eqnarray}

Therefore, the first law of black holes in CEPYM theory, eq.(\ref{1stlaw}), can be rewritten as
\begin{eqnarray}
&&\delta\mathcal{E}=\Omega_H\delta\mathcal{J}+
\frac{\kappa}{2\pi}\delta S
-\int_{S_H}A_{\beta}^a\xi^{\beta}\delta(\epsilon_{\alpha_3\alpha_4\alpha\mu}F^{\alpha\mu}_a).\ \ \ \ \
\end{eqnarray}

For the simple scalar background $\phi=1$, eq.(\ref{1stlaw}) reduces to 
\begin{eqnarray}
\delta\mathcal{E}=\Omega_H\delta\mathcal{J}+
\frac{1}{6}\kappa\delta A
-\int_{S_H}A_{\beta}^a\xi^{\beta}\delta(\epsilon_{\alpha_3\alpha_4\alpha\mu}F^{\alpha\mu}_a).
\end{eqnarray}
which is consistent with the result obtained in \cite{Gao2003, HS1993}.

\section{Conclusion}\label{conclusion}

In this work, we have systematically investigated the thermodynamic properties of black holes in the conformal Einstein-power-Yang–Mills theory within the covariant phase space formalism. We first derived the explicit expressions of the symplectic potential 3-form and decomposed the Noether current 3-form into a constraint 3-form and the exterior derivative of the Noether charge 2-form, thereby clarifying the conserved charge structure of the CEPYM field system. Through  asymptotic expansion and near-horizon geometric analysis, we have established the generalized first law of black hole mechanics for stationary axisymmetric black holes in CEPYM theory. The derived thermodynamic formula incorporates corrections arising from the conformal scalar field and non-Abelian Yang–Mills gauge field, thereby generalizing the classic black hole first law. The non-trivial gauge contribution originates from the intrinsic nonlinearity of the $SU(2)$ Lie algebra, distinct from that of linear electromagnetic fields. This work extends black hole thermodynamics to conformally coupled gravity-gauge-scalar systems and provides a theoretical basis for further exploration of thermodynamic stability and phase transition behaviors of generalized black hole solutions.

\section*{Acknowledgments}
We thank Dr. Jie Jiang for helpful discussions. X. He is partially supported by the National Natural Science Foundation of
China (12475049). X. Wu is partially supported by the National Natural
Science Foundation of China (12275350). N. Xie is partially sponsored by
the Natural Science Foundation of Shanghai (24ZR1406000).

\appendix


\end{document}